\documentclass[sigconf, screen, authorversion]{acmart} 

\usepackage{enumitem}
\usepackage{microtype}
\usepackage{subfigure}
\usepackage{booktabs} 
\usepackage{hyperref}
\usepackage{latexsym}
\RequirePackage{fix-cm}
\usepackage{amsmath, mathtools}
\usepackage{relsize}
\usepackage{footnote}
\usepackage[rightcaption]{sidecap}
\makesavenoteenv{tabular}
\makesavenoteenv{table}
\usepackage{multirow}
\usepackage{array, url}
\usepackage{dblfloatfix}
\usepackage{graphicx}
\DeclareGraphicsExtensions{.pdf,.jpg,.png}
\graphicspath{{figures/}}
\usepackage{wrapfig}

\AtBeginDocument{%
  \providecommand\BibTeX{{%
    \normalfont B\kern-0.5em{\scshape i\kern-0.25em b}\kern-0.8em\TeX}}}

\copyrightyear{2022}
\acmYear{2022}
\setcopyright{rightsretained}
\acmConference[FAccT '22]{2022 ACM Conference on Fairness, Accountability, and Transparency}{June 21--24, 2022}{Seoul, Republic of Korea}
\acmBooktitle{2022 ACM Conference on Fairness, Accountability, and Transparency (FAccT '22), June 21--24, 2022, Seoul, Republic of Korea}\acmDOI{10.1145/3531146.3533089}
\acmISBN{978-1-4503-9352-2/22/06}

\begin{document}

\title{Bias in Automated Speaker Recognition}

\author{Wiebke Toussaint Hutiri}
\authornote{Corresponding author}
\email{w.toussaint@tudelft.nl}
\orcid{}
\author{Aaron Yi Ding}
\orcid{}
\affiliation{%
  \institution{Delft University of Technology}
  \streetaddress{Jaffalaan 5, 2628 BX}
  \city{Delft}
  \country{The Netherlands}
}

\renewcommand{\shortauthors}{Hutiri and Ding}

\begin{abstract}
Automated speaker recognition uses data processing to identify speakers by their voice. Today, automated speaker recognition is deployed on billions of smart devices and in services such as call centres. Despite their wide-scale deployment and known sources of bias in related domains like face recognition and natural language processing, bias in automated speaker recognition has not been studied systematically. We present an in-depth empirical and analytical study of bias in the machine learning development workflow of speaker verification, a voice biometric and core task in automated speaker recognition. Drawing on an established framework for understanding sources of harm in machine learning, we show that bias exists at every development stage in the well-known VoxCeleb Speaker Recognition Challenge, including data generation, model building, and implementation. Most affected are female speakers and non-US nationalities, who experience significant performance degradation. Leveraging the insights from our findings, we make practical recommendations for mitigating bias in automated speaker recognition, and outline future research directions.

\end{abstract}

\ccsdesc[500]{General and reference~Evaluation}
\ccsdesc[500]{Security and privacy~Biometrics}
\ccsdesc[300]{Computing methodologies~Speech recognition}
\ccsdesc[100]{Computing methodologies~Machine learning}

\keywords{speaker recognition, speaker verification, bias, fairness, audit, evaluation}

\maketitle

\section{Introduction}
\label{introduction}


The human voice contains an uncanny amount of personal information. Decades of research have correlated behavioural, demographic, physiological, sociological and many other individual characteristics to a person's voice. Even if untrained human listeners cannot discern all the details, automated voice processing can: voice profiling can reveal sensitive personal attributes such as age, anatomy, health status, medical conditions, identity, intoxication, emotional state, stress, and truthfulness from speech~\cite{Singh2019Profiling}. Speaker recognition is a type of voice processing that automatically recognises the identity of a human speaker from personal information contained in their voice~\cite{Furui1994Overview}. Today, speaker recognition permeates private and public life. Speaker recognition systems are deployed at scale in call centers, on billions of mobile phones and on voice-enabled consumer devices such as smart speakers. They grant access not only to personal devices in intimate moments, but also to essential public services for vulnerable user groups. For example, in Mexico speaker recognition is used to allow senior citizens to provide a telephonic proof-of-life to receive their pension~\cite{veridas2021}. 

In this paper we study bias in speaker recognition systems. Bias in machine learning (ML) is a source of unfairness~\cite{mehrabi2019survey} that can have harmful consequences, such as discrimination~\cite{wachter2021bias}. Bias and discrimination in the development of face recognition technologies~\cite{buolamwini2018gendershades, Raji2019actionable, Raji2021aboutface}, natural language processing~\cite{Bolukbasi2016Man} and automated speech recognition~\cite{addadecker2005do, tatman2017effects, koenecke2020racial, Toussaint2021Characterising} are well studied and documented. Bias in speaker recognition, a related domain, has received very limited attention. Yet, speaker recognition technologies are pervasive and process extremely sensitive personal data that is intricately intertwined with our individual identity. They are deployed in high-stakes applications, while the modality of their input data makes them susceptible to perpetrate discrimination. It is thus urgent to investigate bias in these systems, so that mitigating and regulatory actions can be taken to forestall potential negative consequences.


Drawing on Suresh and Guttag's \emph{Framework for Understanding Sources of Harm}~\cite{Suresh2021Framework}, we present the first detailed study on bias in speaker recognition. We approach this work as a combination of an analytical and empirical evaluation focused on the VoxCeleb Speaker Recognition Challenge~\cite{Nagrani2020Voxsrc}, one of the most popular benchmarks in the domain with widely used datasets. Our study shows that existing benchmark datasets, learning mechanisms, evaluation practices, aggregation habits and post-processing choices in the speaker recognition domain produce systems that are biased against female and non-US speakers. 
Our contributions are:
\begin{enumerate}
    \item We present an evaluation framework for quantifying performance disparities in speaker verification - a speaker recognition task that serves as the biometrics of voice
    \item We apply this framework to conduct the first evaluation of bias in speaker verification. Our results show that bias exists at every stage of the ML development pipeline
    \item Informed by our evaluation, we recommend research directions to address bias in automated speaker recognition
\end{enumerate}

Our paper is structured as follows. In Section~\ref{s:related_work} and~\ref{s:background} we review related work and provide a background on speaker recognition, its evaluation and supporting infrastructure for its development. We then present the empirical experiment setup and bias evaluation framework in Section~\ref{s:method}. In Section~\ref{s:data_generation_bias} we present our findings of bias in data generation, and in Section~\ref{s:model_building_implementation_bias} our findings of bias in model building and implementation. We discuss our findings, and make recommendations for mitigating bias in speaker recognition in Section~\ref{s:discussion}. Finally, we conclude in Section~\ref{s:conclusion}.

\section{Related Work}
\label{s:related_work}

In this section we provide a background on speaker recognition within its historical development context and present evidence of bias in the domain. We then introduce the theoretical framework on which we base our analytical and empirical bias evaluation.

\subsection{Historical Development of Automated Speaker Recognition}

Since its inception, research into speaker recognition has enabled voice-based applications for access control, transaction authentication, forensics, law enforcement, speech data management, personalisation and many others~\cite{Reynolds2002Overview}. As a voice-based biometric, speaker verification is viewed to have several advantages: it is physically non-intrusive, system users have historically considered it to be non-threatening, microphone sensors are ubiquitously available in telephone and mobile systems or can be installed at low-cost if they are not, and in many remote applications speech is the only form of biometrics available~\cite{Reynolds2002Overview}. Given the proliferation of speaker recognition systems in digital surveillance technologies, concerns over its pervasive, hidden and invasive nature are rising~\cite{eff2021catalog}. 

\subsubsection{Parallels to Facial Recognition}
The historical development of automated speaker recognition reflects that of facial recognition in many aspects. Similar to the development of facial recognition systems~\cite{Raji2021aboutface}, research in early speaker recognition systems was supported by defense agencies, with envisioned applications in national security domains such as forensics~\cite{greenberg2020two}. The systems relied on datasets constructed from telephone corpuses and their development was greatly accelerated through coordinated, regular competitions and benchmarks. 

\subsubsection{From Classical Approaches to Deep Neural Networks}
Two years after the deep learning breakthroughs in computer vision, Deep Neural Networks (DNNs) were first applied to speaker recognition systems~\cite{heigold2016endtoend}. Since 2016, DNNs have become the dominant technique for developing speaker recognition systems~\cite{snyder2017deep, li2017deep, snyder2018xvectors}. DNNs have distinguished themselves in important ways from traditional approaches for speaker recognition: their performance is superior on short speech utterances~\cite{snyder2018xvectors}, they can be trained in an end-to-end fashion using only speaker labels, thus reducing laborious labelling efforts~\cite{heigold2016endtoend}, and they can leverage many of the techniques that have demonstrated success in the image recognition domain. To enable the new era of deep speaker recognition, large scale datasets were needed to support research in this emerging area, and methods for generating them adapted approaches from face recognition. For example, a popular speaker recognition dataset, VoxCeleb~\cite{Nagrani2017Voxceleb}, is derived from the voice signals in Youtube videos of celebrities contained in the well-known face recognition dataset VGG Face~\cite{Parkhi2015Deep}. Another dataset, MOBIO~\cite{Khoury2014bimodal}, was developed jointly for mobile face recognition and speaker recognition.

\subsection{Bias in Speaker and Speech Recognition}

\subsubsection{Early Evidence of Bias in Speaker Recognition}
It is well established that speaker characteristics such as age, accent and gender affect the performance of speaker recognition~\cite{hansen2015speaker}. In acknowledgement of this, past works in speech science, like research promoted through the 2013 Speaker Recognition Evaluation in Mobile Environments challenge, have reported speaker recognition performance separately for male and female speakers~\cite{Khoury2013speaker}. The submissions to the challenge made it clear that bias is a cause of concerns: of 12 submissions, all submitted systems performed worse for females than for males on the evaluation set. On average the error rate for females was 49.35\% greater than for males. Despite these performance differences being acknowledged, they went unquestioned and were attributed solely to an unbalanced training set that contained a male:female speaker ratio of 2:1. In later works the discrepancy between female and male speakers is still evident and reported, but remains unquestioned and unaddressed~\cite{Park2016Speaker}. Historically, a common approach to avoid gender-based bias has been to develop separate models for female and male speakers~\cite{Kinnunen2009Overview}. While this may be insufficient to eradicate bias, generating separate feature sets for female and male speakers can reduce it~\cite{Mazairafernandez2015improving}. Beyond considering binary gender, evaluating demographic performance gaps based on other speaker attributes is less common, and intersectional speaker subgroups have not been considered.


\subsubsection{Nuanced Evaluation No Longer Common Practice}
Since the adoption of Deep Neural Networks (DNNs) for speaker recognition, practices of evaluating system performance for speaker subgroups seem to have disappeared. Several system properties beyond performance have been considered in recent years, such as robustness~\cite{Bai2021Speaker} and privacy~\cite{Nautsch2019preserving}. However, research in robustness and privacy in speaker recognition does not address the glaring gap that remains in the domain: system performance appears biased against speaker groups based on their demographic attributes. Only one recent study investigates bias in end-to-end deep learning models based on speaker age and gender~\cite{Fenu2021Fair}, reconfirming the importance of balanced training sets. 

\subsubsection{Bias in Automated Speech Recognition}
In automated speech recognition, which is concerned with the linguistic content of voice data, not with speaker identity, recent studies have provided evidence that commercial automated caption systems have a higher word error rate for speakers of colour \cite{tatman2017effects}. Similar racial disparities exist in commercial speech-to-text systems, which are strongly influenced by pronunciation and dialect \cite{koenecke2020racial}. Considering their shared technical backbone with facial recognition systems, and shared data input with automated speech recognition systems, we expect that bias and harms identified in these domains will also exist in speaker recognition systems. Mounting evidence of bias in facial and speech recognition, the abundance of historic evidence of bias and the vacuum of public information about bias in speaker recognition, strengthen the motivation for our work.

\subsection{Sources of Harm in the ML Life Cycle}
\label{ss:sources_of_harm}
We draw on Suresh and Guttag's~\cite{Suresh2021Framework} \emph{Framework for Understanding Sources of Harm} through the ML life cycle to ground our investigation into bias in automated speaker recognition. Suresh and Guttag divide the ML life cycle into two streams and identify seven sources of bias related harms across the two streams: 1) the data generation stream can contain historical, representational and measurement bias; and 2) the model building and implementation stream can contain learning, aggregation, evaluation and deployment bias. \emph{Historical bias} replicates bias, like stereotypes, that are present in the world as is or was. \emph{Representation bias} underrepresents a subset of the population in the sample, resulting in poor generalization for that subset. \emph{Measurement bias} occurs in the process of designing features and labels to use in the prediction problem. \emph{Aggregation bias} arises when data contains underlying groups that should be treated separately, but that are instead subjected to uniform treatment. \emph{Learning bias} concerns modeling choices and their effect on amplifying performance disparities across samples. \emph{Evaluation bias} is attributed to a benchmark population that is not representative of the user population, and to evaluation metrics that provide an oversimplified view of model performance. Finally, \emph{deployment bias} arises when the application context and usage environment do not match the problem space as it was conceptualised during model development. 

Next we introduce automated speaker recognition, and then show analytically and empirically how these seven types of bias manifest in the speaker recognition development ecosystem.

\section{Background}
\label{s:background}

Speaker recognition refers to the collection of data processing tasks that identify a speaker by their voice~\cite{Furui1994Overview}. Core tasks in speaker recognition are \emph{speaker identification}, which determines a speaker's identity from a subset of speakers, \emph{speaker verification}, which validates if a speaker's identity matches the identity of a stored speech utterance, and \emph{speaker diarisation}, which is concerned with partitioning speech to distinguish between different speakers~\cite{Bai2021Speaker}. While technical implementation details differ in the three areas, their communities overlap, they share datasets and participate in the same competitions. We focus our investigation in this paper on speaker verification, which underlies voice biometrics. However, as the tasks have evolved together, many of the biases that we uncover in speaker verification also apply to speaker identification and diarisation. In this section we provide a high level overview of speaker verification and its evaluation, as well as its supporting ecosystem of competitions and benchmarks that have advanced the field. We refer the reader to~\cite{Bai2021Speaker} for a detailed technical survey on state-of-the-art speaker recognition, and to \cite{Kinnunen2009Overview} for a review on the classical speaker recognition literature prior to the advent of Deep Neural Networks (DNNs).


\subsection{Speaker Verification Overview}

A speaker verification system determines whether a candidate speaker matches the identity of a registered speaker by comparing a candidate speaker's speech signal (i.e. \emph{trial utterance}) to the speech signal of a registered speaker (i.e. \emph{enrollment utterance}). Speaker verification is classified based on its training data as text-dependent if speech signals are fixed phrases or text-independent if not, prompted if speech was produced by reading text or spontaneous if not~\cite{greenberg2020two}. Spontaneous text-independent speech is the type of speech that occurs naturally when a speaker interacts with a voice assistant or a call centre agent, and presents the most general speaker verification task. 

\begin{figure}[t]
    \centering
    \includegraphics[width=\linewidth]{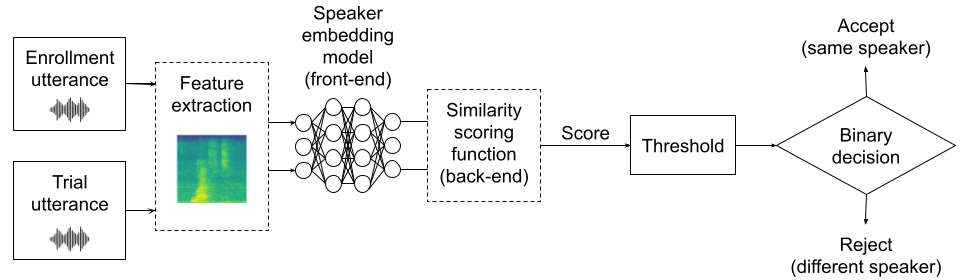}
    \caption{Speaker verification data processing pipeline}
    \label{fig:speaker_verification_overview}
\end{figure} 

As shown in Figure~\ref{fig:speaker_verification_overview}, many speaker verification systems consists of two stages, a front-end that generates a speaker embedding model for enrollment and trial utterances, and a back-end that computes a similarity score for the two resultant embeddings. Alternatively, end-to-end speaker verification directly learns a similarity score from training utterances~\cite{heigold2016endtoend}. Modern speaker verification systems use DNNs to learn the front-end embedding, or to train the end-to-end system~\cite{Bai2021Speaker}. As the final step of the speaker verification process, the score output is compared to a threshold. Speaker identity is accepted if the score lies above the threshold, and rejected if it lies below the threshold. 


\subsection{Speaker Verification Evaluation}
\label{ss:speaker_verification_evaluation}

To evaluate speaker verification systems, scores are generated for many pairs of enrollment and trial utterances. The utterance pairs are labelled as being from the same or from different speakers. Two typical score distributions generated from many same and different speaker utterance pairs are shown in Figure~\ref{fig:scores_example} in the Appendix. After calibrating the speaker verification system to a threshold (e.g. \emph{equal error rate} or \emph{detection cost}), utterance pairs with a score below the threshold are classified as different speakers and the trail utterance is rejected. Utterance pairs with a score above the threshold are classified as the same speaker, and accepted. As the two distributions overlap, classification is not perfect. At a particular threshold value there will be false positives, i.e. utterance pairs of different speakers with a score above the threshold, and false negatives, i.e. utterance pairs of the same speakers with a score below the threshold. 


\noindent
Speaker verification performance is determined by its false positive rate (FPR) and false negative rate (FNR) at the threshold value to which the system has been calibrated~\cite{greenberg2020two}. It is accepted that the two error rates present a trade-off, and that selecting an appropriate threshold is an application-specific design decision~\cite{NIST2020}. The threshold value is determined by balancing the FPR and FNR error rates for a particular objective, such as obtaining an \emph{equal error rate} (EER) for FPR and FNR, or minimising a cost function. The \emph{detection cost function} (DCF) is a weighted sum of FPR and FNR across threshold values, with weights determined by the application requirements. To compare performance across models, systems are frequently tuned to the threshold value at the minimum of the DCF, and the corresponding \emph{detection cost} $C_{Det}$ value is reported as a metric. Various detection cost functions have been proposed over time, such as the following, proposed in the NIST SRE 2019 Evaluation Plan~\cite{NIST2019}:

\begin{equation}
\small
\begin{split}
    C_{Det}\left(\theta\right) & = C_{FN} \times P_{Target} \times P_{FN}\left(\theta\right) + C_{FP} \times \left(1 - P_{Target}\right) \times P_{FP}\left(\theta\right) \\
    & P_{Target} = 0.05, \;
    C_{FN} = 1, \;
    C_{FP} = 1    
\label{eq:cdet}
\end{split}
\vspace{-2mm}
\end{equation}

\noindent
Speech science literature recommends that \emph{detection error trade-off} (DET) curves~\cite{greenberg2020two} are used to visualise the trade-off between FPR and FNR, and to consider system performance across various thresholds. DET curves visualise the FPR and FNR at different operating thresholds on the x- and y-axis of a normal deviate scale \cite{Martin1997Det} (see Figure~\ref{fig:det_curve_example} in the Appendix). They can be used to analyse the inter-model performance (across models), and are also recommended for analysing intra-model performance (across speaker subgroups in a model).

\begin{table*}[h]
    \small
    \centering
    \renewcommand{\arraystretch}{1.2}
    \begin{tabular}{l|l|c|c}
        \textbf{Name} & \textbf{Organiser} & \textbf{Years} & \textbf{Metrics} \\
        NIST SRE \cite{greenberg2020two} & US National Inst. of Standards \& Tech. & 1996 - 2021 & Detection Cost Function\\
        SRE in Mobile Env's \cite{Khoury2013speaker} & Idiap Research Institute & 2013 & DET curve, EER, \emph{half total error rate}\\
        Speakers in the Wild SRC \cite{mclaren2016speakers} & at Interspeech 2016 & 2016 & $min\ C_{Det}$* (SRE2016), $R_{prec}$, $C_{llr}$\\
        VoxCeleb SRC \cite{Nagrani2020Voxsrc} & Oxford Visual Geometry Group & 2019 - 2021 &  $min\ C_{Det}$* (SRE2018), EER\\
        Far-Field SVC \cite{qin2020interspeech} & at Interspeech 2020 & 2020 & $min\ C_{Det}$*, EER \\
        Short Duration SVC \cite{zeinali2019shortduration} & at Interspeech 2021 & 2020 - 2021 & $norm\ min\ C_{Det}$* (SRE08)\\
        SUPERB benchmark \cite{yang2021superb} & CMU, JHU, MIT, NTU, Facebook AI & 2021 & EER* \\
        \end{tabular} {\medskip}
    \caption{Evaluation metrics for Speaker Verification and Recognition Challenges (SVC and SRC) (* primary metric)}
    \label{tab:sv_challenges}
\vspace{-5mm}
\end{table*}

\subsection{Competitions and Benchmarks}

Speaker recognition challenges have played an important role in evaluating and benchmarking advances in speaker verification. They were first initiated within the Information Technology Laboratory of the US National Institute of Standards and Technology (NIST) to conduct evaluation driven research on automated speaker recognition~\cite{greenberg2020two}. The NIST Speaker Recognition Evaluation (SRE) challenges and their associated evaluation plans have been important drivers of speaker verification evaluation. In addition, new challenges have emerged over time to address the requirements of emerging applications and tasks. Table \ref{tab:sv_challenges} summarises recent challenges, their organisers and the metrics used for evaluation. Most challenges have adopted the minimum of the detection cost function, $min\ C_{Det}$, recommended by the NIST SREs as their primary metric. As the NIST SREs have modified this function over time, different challenges use different versions of the metric. In the remainder of this paper we evaluate bias in the VoxCeleb Speaker Recognition Challenge (SRC).
\section{Experiment Setup}
\label{s:method}

Launched in 2019, the objective of the VoxCeleb SRC is to "probe how well current [speaker recognition] methods can recognise speakers from speech obtained `in the wild'"~\cite{voxcelebsrc2021}. The challenge has four tracks: open speaker diarisation, open and closed fully supervised, and closed self-supervised speaker verification. It serves as a well-known benchmark, and has received several hundred submissions over the past three years. The popularity of the challenge and its datasets make it a suitable candidate for our evaluation, representative of the current ecosystem. 
We evaluate group bias in the speaker verification track of the VoxCeleb SRC. 

\subsection{Baseline Models}

The challenge has released two pre-trained baseline models~\cite{heo2020clova} trained on the VoxCeleb~2 training set~\cite{Nagrani2020a} with close to 1 million speech utterances of 5994 speakers. 61\% of speakers are male and 29\% of speakers have a US nationality, which is the most represented nationality. More detailed metadata is not readily available. The baseline models are based on a 34-layer ResNet trunk architecture. \emph{ResNetSE34V2}~\cite{heo2020clova} is a larger model, with an architecture optimised for predictive performance. \emph{ResNetSE34L}~\cite{chung2020defence} is a smaller model that contains less than a fifth of the parameters of \emph{ResNetSE34V2} and has smaller input dimensions. This reduces the computation time and the memory footprint of the model, two important considerations for on-device deployment in applications like smartphones and smart speakers. The model developers have optimised it for fast execution. We downloaded and used both baseline models as black-box predictors in our evaluation. The technical details of the baseline models are summarised in Table~\ref{tab:baseline_model_attributes} in the Appendix.


\subsection{Evaluation Dataset}
\label{ss:evaluation_dataset}
We evaluate the baseline models on three established evaluation sets that can be constructed from the utterances in the VoxCeleb~1 dataset~\cite{Nagrani2020a}. VoxCeleb~1 was released in 2017 with the goal of creating a large scale, text-independent speaker recognition dataset that mimics unconstrained, real-world speech conditions, in order to explore the use of DNNs for speaker recognition tasks \cite{Nagrani2017Voxceleb}. The dataset contains 153 516 short clips of audio-visual utterances of 1251 celebrities in challenging acoustic environments (e.g. background chatter, laughter, speech overlap) extracted from YouTube videos. The dataset also includes metadata for speakers' gender and nationality, and is disjoint from VoxCeleb~2 which is used for training. Three different evaluation sets have been designed for testing speaker verification with VoxCeleb~1. We consider all three evaluation sets in our analysis. The evaluation sets are discussed in detail in~\S\ref{ss:evaluationbias}.

\subsection{Speaker Subgroups and Bias Evaluation Measures}
\label{ss:bias_evaluation_measure}
We selected subgroups based on attributes and categories captured in the VoxCeleb metadata: gender and nationality.
We then established bias by evaluating performance disparities between these subgroups using existing evaluation measures in speaker verification. Reusing attributes and category labels, though practical for facilitating our study, perpetuates existing bias. We reflect on the consequences of this in our analysis of measurement bias in \S\ref{ss:measurement_bias}.

Our first technique for establishing bias is to plot the DET curves for all subgroups, and to compare the subgroups' DET curves to the overall curve for all subgroups. As speaker verification systems must operate on the DET curve, this presents the theoretical performance boundary of the model across subgroups. Secondly, we consider bias at the threshold to which the system has been calibrated, which ultimately presents the operating point of the system. Here we consider an unbiased system as one that has equal false positive and true positive (or false negative) rates across subgroups, in line with the definition of equalized odds~\cite{Hardt2016equality}. We compare each subgroup's performance to the overall system performance to facilitate comparison across a large number of subgroups, and thus deviate slightly from the formal definition of equalized odds. We use $C_{Det}(\theta)$ as defined in Equation~\ref{eq:cdet} to determine the calibration threshold and quantify the relative bias towards each subgroup with the ratio of the subgroup cost $C_{Det}\left(\theta\right)^{SG}$ to the overall cost  $C_{Det}\left(\theta\right)^{overall}$ at the threshold value where $C_{Det}(\theta)$ is minimized for the overall system: 

\begin{equation}
    subgroup\ bias = \frac{C_{Det}\left(\theta_{@\ overall\ min}\right)^{SG}}{C_{Det}\left(\theta_{@\ overall\ min}\right)^{overall}}
    \label{eq:sgbias}
\end{equation}

\noindent
If the \emph{subgroup bias} is greater than 1, the subgroup performance is worse than the overall performance, and the speaker verification model is prejudiced against that subgroup. Conversely, if the \emph{subgroup bias} is less than 1, the model favours the subgroup. If the ratio is exactly 1, the model is unbiased for that subgroup.




\subsection{Black-box Bias Evaluation Framework}
\begin{figure}[bth]
    \centering
    \includegraphics[width=\linewidth]{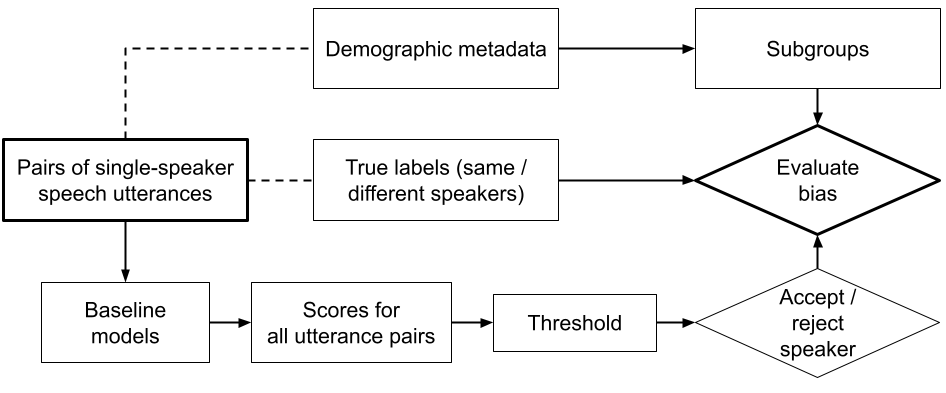}
    \caption{Framework for black-box bias evaluation of speaker verification models}
    \label{fig:evaluation_framework}
    \vspace{-2mm}
\end{figure}

We designed a framework\footnote{The code for the evaluation has been released as an open-source python library: \url{https://github.com/wiebket/bt4vt/releases/tag/v0.1}} that replicates a real evaluation scenario to evaluate bias in the VoxCeleb SRC benchmark. Figure~\ref{fig:evaluation_framework} shows an overview of our approach. We start with pairs of single-speaker speech utterances in the evaluation dataset as input, and use the baseline models, \emph{ResNetSE32V2} and \emph{ResNetSE34L}, as black-box predictors. The baseline models output scores for all utterance pairs in the evaluation set. We set the threshold to the value that minimizes the overall system cost of the DCF and accept or reject speakers in utterance pairs based on that. Our predicted binary acceptance is then compared to the true labels of the utterance pairs to determine false positive and false negative predictions. Using the metadata for speakers, we allocate each utterance pair to a subgroup based on the attributes of the enrollment utterance. From these inputs we evaluate bias by establishing the FPR, FNR and thus $C_{Det}\left(\theta\right)^{SG}$ at the threshold value for each subgroup. We also plot DET curves from the outputs scores for each subgroup. The evaluation is repeated for each of the three VoxCeleb~1 evaluation sets. Using this evaluation framework, we now identify sources of bias in data generation (Section~\ref{s:data_generation_bias}) and model building and implementation (Section~\ref{s:model_building_implementation_bias}).

\section{Bias in Data Generation}
\label{s:data_generation_bias}

In this section we identify sources of bias in the VoxCeleb SRC that arise during data generation (see Suresh and Guttag's \emph{Framework for Understanding Sources of Harm} described in \S\ref{ss:sources_of_harm}). The stage involves data generation, population definition and sampling, measurement and pre-processing, with the goal of creating training, test and benchmark datasets. The types of bias that arise in these processes are \emph{historical bias}, \emph{representation bias} and \emph{measurement bias}. 

\subsection{Historical Bias}
\emph{Historical bias replicates biases, like stereotypes, that are present in the world as is or was.}

\smallskip
\noindent
The VoxCeleb~1 dataset was constructed with a fully automated data processing pipeline from open-source audio-visual media~\cite{Nagrani2017Voxceleb}. The candidate speakers for the dataset were sourced from the VGG Face dataset~\cite{Parkhi2015Deep}, which is based on the intersection of the most searched names in the Freebase knowledge graph and Internet Movie Database (IMDB). After searching and downloading video clips for identified celebrities, further processing was done to track faces, identify active speakers and verify the speaker's identity using the HOG-based face detector~\cite{King2009Dlibml}, Sync-Net~\cite{Chung2017out} and VGG Face CNN~\cite{Simonyan2015very} respectively. If the face of a speaker was correctly identified, the clip was included in the dataset. 

Bias in facial recognition technologies is well known~\cite{buolamwini2018gendershades, Raji2019actionable, Raji2021aboutface}, and historic bias pervades the automated data generation process of VoxCeleb. The VoxCeleb~1 inclusion criteria subject the dataset to the same bias that has been exposed in facial recognition verification technology and reinforce popularity bias based on search results~\cite{mehrabi2019survey}. Moreover, the data processing pipeline directly translates bias in facial recognition systems into the speaker verification domain, as failures in the former will result in speaker exclusion from VoxCeleb~1.

\subsection{Representation Bias}
\label{ss:representation_bias}
\emph{Representation bias underrepresents a subset of the population in its sample, resulting in poor generalization for that subset.}

\smallskip
\noindent
The VoxCeleb~1 dataset is skewed towards males and US nationals, as can be seen in Figure~\ref{fig:speaker_demographics} in the Appendix. Performance for this group is the most reliable and aligns the closest with the average performance. For subgroups with the smallest amount of speakers, such as Italian, German and Irish females, DET curves in Figure~\ref{fig:resnetse34v22_det_curves_nationality_all} in the Appendix show that performance is unreliable. In the context of benchmark evaluations, such skewed representation not only provides an inadequate understanding of the real capabilities of speaker verification for a diverse population of people, but it also shapes the development of the technology towards the group of people that are most represented. Representation bias affects the quality of our bias evaluation for underrepresented subgroups. However, there are sufficient subgroups that have a reasonable representation of speakers (USA, Canadian, UK, Indian and Australian males and females) to support our efforts of gathering evidence of bias. 

Recent work on age recognition with the VoxCeleb datasets~\cite{Hechmi2021voxceleb} shows that speakers between ages 20 and 50 are most represented in the dataset, indicating that representation bias is also evident across speaker age. Nationality, gender and age only account for some of the attributes of a speaker's voice that affect automated speaker recognition~\cite{Singh2019Profiling}. We discuss additional attributes that are likely to affect performance in the following section on \emph{measurement bias}. Being a celebrity dataset that is not representative of the broad public, it is likely that VoxCeleb~1 contains representation bias that affect many other sensitive speaker attributes. Representation bias contributes to aggregation bias (\S\ref{ss:aggregationbias}), evaluation bias (\ref{ss:evaluationbias}) and deployment bias (\S\ref{ss:deploymentbias}) in speaker verification, and is discussed in further detail in those sections.

\subsection{Measurement Bias}
\label{ss:measurement_bias}
\emph{Measurement bias occurs in the process of designing features and labels to use in the prediction problem.}

\smallskip
\noindent
In our analysis of measurement bias we focus on labelling choices made in the VoxCeleb~1 metadata, which our study inherits in our subgroup design choices. While these labels are not used for making predictions, they are used to make judgements about speaker representation in the dataset. They also inform subgroup design, which plays a fundamental role in our group-based bias analysis.

The VoxCeleb~1 dataset creators inferred nationality labels from speakers' countries of citizenship, as obtained from Wikipedia. Their underlying motivation for doing this was to assign a label that is indicative of a speaker's accent~\cite{Nagrani2020a}. Conflating nationality and accent is problematic, as people with the same citizenship can speak the same language with different accents. Likewise, many countries have citizens speaking different languages (e.g. India has 7 languages with more than 50~million first language speakers each~\cite{wiki:List_of_languages_by_number_of_native_speakers_in_India}).
Using nationality as a subgroup label has merits\footnote{Discrimination based on national origin can have legal consequences, for instance, Title VII of the Civil Rights Act of 1964 prohibits employment discrimination based on national origin in the United States}, even if conflating nationality, accent and language raises concerns. 
The nationality-based performance differences that we observe suggest that language, accent, ethnicity and dialect may also produce disparate performance.


The metadata considers only binary gender categories, namely male and female. From the dataset description it is unclear what method was followed to label speakers by gender. Many concerns about gender labelling in face analysis technologies have been pointed out in prior research~\cite{Scheuerman2019how}, and similar concerns hold true in speaker recognition. Simply replacing a binary gender classification with more categories is not a recommended alternative. Even if it were possible to produce accurate labels, they might help to mitigate bias in speaker verification only while offering a new surface for harm, for example through voice-based gender classification enabled targeting. 


\section{Bias in Model Building and Implementation}
\label{s:model_building_implementation_bias}

Having analysed bias in data generation, we now present evidence of bias in the model building and implementation stage of the VoxCeleb SRC benchmark. 
In the ML pipeline this stage involves model definition and training, evaluation and real-world deployment. The types of bias that arise in these processes are \emph{aggregation bias}, \emph{learning bias}, \emph{evaluation bias} and \emph{deployment bias}. We found evidence of each type of bias in our evaluation. 

\subsection{Aggregation Bias}
\label{ss:aggregationbias}
\emph{Aggregation bias arises when data contains underlying groups that should be treated separately, but that are instead subjected to uniform treatment.}

\begin{figure*}[hbt]
    \centering
    \includegraphics[width=0.9\textwidth]{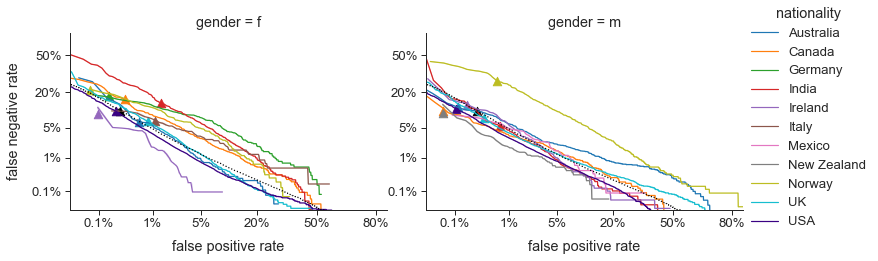}
    \vspace{-3mm}
    \caption{Aggregation bias in \emph{ResNetSE34V2} with the \emph{VoxCeleb~1-H} evaluation set: 1) the aggregate model (dotted black line) is fit to the dominant population (US speakers) and 2) the operating performance (triangular markers) across subgroups has high variability when the system is tuned to the overall threshold.}
    \label{fig:resnetse34v22_det_curves_gender}
    \vspace{-3mm}
\end{figure*}

\smallskip
\noindent
We evaluate aggregation bias by plotting disaggregated DET performance curves for speaker subgroups based on nationality and gender. In Figure \ref{fig:resnetse34v22_det_curves_gender} we show the DET curves for female (left) and male (right) speakers across 11 nationalities for the \emph{ResNetSE34V2} model evaluated on the \emph{VoxCeleb~1-H} evaluation set. The dotted black DET curve shows the overall performance across all subgroups. DET curves above the dotted line have a high likelihood of performing worse than average, while DET curves below the dotted line will generally perform better than average. It is easy to see that the DET curves of female speakers lie mostly above the average DET curve, while those of male speakers lie below it. The model is thus likely to perform worse than average for females, and better for males. Figure~\ref{fig:resnetse34v22_det_curves_nationality_all} in the Appendix shows DET subplots for each nationality, highlighting disparate performance across nationalities.

The triangular markers show the FPR and FNR at the threshold $\theta_{@\ overall\ min}$ where the overall system DCF is minimized. The markers for male and female speaker subgroups are dispersed, indicating that the aggregate system calibration results in significant operating performance variability across subgroups. Table~\ref{tab:cdet_ratios_resnetse34v22} in the Appendix shows the $subgroup\ bias$ for all subgroups. With the exception of US female speakers, all females have a $subgroup\ bias$ greater than 1, and thus perform worse than average. 

The DET curves and $subgroup\ bias$ demonstrate disparate performance based on speakers' gender and nationality. They also show that the model is fit to the dominant population in the training data, US speakers.
The trends in aggregation bias that we observe for \emph{ResNetSE34V2} are evident in all three evaluation sets, as well as \emph{ResNetSE34L}. They indicate that speaker verification models do not identify all speaker subgroups equally well, and validate that performance disparities between male and female speakers identified in the past~\cite{Mazairafernandez2015improving} still exist in DNN speaker verification models today.

\subsection{Learning Bias}
\label{ss:learningbias}
\emph{Learning bias concerns modeling choices and their effect on amplifying performance disparities across samples.}

\smallskip
\noindent
The \emph{ResNetSE34V2} and \emph{ResNetSE34L} models are built with different architectures and input features. The two architectures have been designed for different goals respectively: to optimise performance and to reduce inference time. Optimisation here refers to the design goal, not the optimiser of the model. The reduced number of parameters, smaller model size and reduced number of computations of \emph{ResNetSE34L} are desirable attributes for on-device deployment. 

\begin{figure}[b]
\vspace{3mm}
    \centering
    \includegraphics[width=\linewidth]{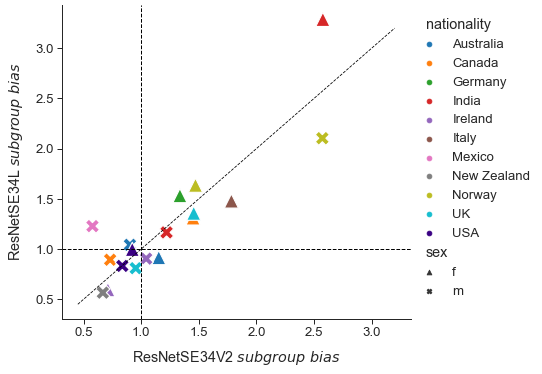}
    \caption{Learning bias based on model architecture. \emph{Subgroup bias} for the performance optimized \emph{ResNetSE34V2} is shown on the x-axis, and for the speed optimized \emph{ResNetSE34L} on the y-axis. On the diagonal \emph{subgroup bias} for the two models is equal.}
    \label{fig:subgroup_bias_models}
\end{figure}

In Figure~\ref{fig:subgroup_bias_models} we plot the $subgroup\ bias$ for both models for all subgroups. On the dotted line the models perform equally well for a subgroup. The greater the distance between a marker and the line, the greater the performance disparity between the models for that subgroup. As described in \S\ref{ss:bias_evaluation_measure}, subgroup performance is worse than average if the $subgroup\ bias$ is greater than 1, and better than average if it is less than 1. We make three observations: Firstly, the $subgroup\ bias$ for both models for male subgroups is close to or less than 1, indicating that at the threshold value males experience better than average performance for both models. Secondly, we observe that $subgroup\ bias$ for male US speakers is equal for both models, indicating that performance disparities remain consistent for the over-represented group. Thirdly, we observe that neither of the two models reduces performance disparities definitively: \emph{ResNetSE34V2} has a lower $subgroup\ bias$ for 7 subgroups, \emph{ResNetSE34L} for 10 subgroups.     

In addition to examining $subgroup\ bias$ we have plotted the DET curves for both models across subgroups in Figure~\ref{fig:det_curves_models_nationality} in the Appendix. We observe that the smaller \emph{ResNetSE34L} increases the distance between DET curves for males and females with nationalities from the UK, USA and Ireland, indicating that the model increases performance disparities between male and female speakers of these nationalities. For Australian, Indian and Canadian speakers the distance between DET curves for males and females remains unchanged, while for Norwegian nationalities they lie closer together. Together these results point to learning bias, highlighting that modeling choices such as the architecture, the number of model parameters and the input feature dimensions can amplify performance disparities in speaker verification. The disparities tend to negatively affect female speakers and nationalities with few speakers.  Our results reinforce other studies that have shown that bias can arise when reducing model size during pruning~\cite{Hooker2020Characterising, Toussaint2022Tiny}, but are insufficient to point out the exact modeling choices that affect learning bias. This remains an area of future work.

\subsection{Evaluation Bias}
\label{ss:evaluationbias}
\emph{Evaluation bias is attributed to a benchmark population that is not representative of the user population, and to evaluation metrics that provide an oversimplified view of model performance.}

\subsubsection{Evaluation Datasets} 
Representative benchmark datasets are particularly important during ML development, as benchmarks have disproportionate power to scale bias across applications if models overfit to the data in the benchmark~\cite{Suresh2021Framework}. Three evaluation sets can be constructed from the VoxCeleb~1 dataset to benchmark speaker verification models. \textit{VoxCeleb 1 test} contains utterance pairs of 40 speakers whose name starts with \emph{E}. \textit{VoxCeleb 1-E} includes the \emph{entire} dataset, with utterance pairs sampled randomly. \textit{VoxCeleb 1-H} is considered a \emph{hard} test set, that contains only utterance pairs where speakers have the same gender and nationality. Speakers have only been included in \textit{VoxCeleb 1-H} if there are at least 5 unique speakers with the same gender and nationality. All three evaluation sets contain a balanced count of utterance pairs from same speakers and different speakers. We have calculated the speaker and utterance level demographics for each evaluation set from the dataset's metadata, and summarise the attributes of the evaluation sets in Table~\ref{tab:voxceleb_attributes}.

\begin{table*}[hbt]
\footnotesize
    \centering
    \begin{tabular}{l|ccc}
         & \textbf{VoxCeleb 1 test} & \textbf{VoxCeleb 1-E} & \textbf{VoxCeleb 1-H}\\
        unique speakers & 40 & 1 251 & 1 190 \\
        unique utterance pairs & 37 720 & 579 818  & 550 894 \\
        speaker pairing details & - & random sample & same gender, nationality \\        
        speaker pair inclusion criteria & name starts with 'E' & all & >=5 same gender, nationality speakers \\
        female / male speakers (\%) & 38 / 62 & 45 / 55 & 44 / 56 \\
        female / male utterances (\%) & 29.5 / 70.5 & 41.8 / 58.2 & 41.1 / 58.9 \\
        count of nationalities & 9 & 36 & 11 \\
        top 1 nationality (\% speakers / utterances) & US (62.5 / 59.6) & US (63.9 / 61.4) & US (67.1 / 64.7)\\
        top 2 nationality (\% speakers / utterances) & UK (12.5 / 13.9) & UK (17.2 / 18.3) & UK (18.1 / 19.3) \\
        top 3 nationality (\% speakers / utterances) & Ireland (7.5 / 6.7) & Canada (4.3 / 3.8) & Canada (4.5 / 3.9) \\
    \end{tabular}
    \smallskip
    \caption{VoxCeleb 1 evaluation sets show that the benchmark's population is not representative across gender and nationality}
    \label{tab:voxceleb_attributes}
\vspace{-4mm}
\end{table*}

\noindent
Several observations can be made based on the summary in Table~\ref{tab:voxceleb_attributes}: the VoxCeleb~1 dataset suffers from representation bias (see Section~\ref{ss:representation_bias}) and all three evaluation sets over-represent male speakers and US nationals. Furthermore, the sample size of \emph{VoxCeleb~1 test} is too small to use it for a defensible evaluation. Its inclusion criterion based on speakers' names introduces additional representation bias into the evaluation set, as names strongly correlate with language, culture and ethnicity. 

In addition to these obvious observations, the summary also reveals subtler discrepancies. Speaker verification evaluation is done on utterance pairs. Demographic representation is thus important on the speaker level to ensure that the evaluation set includes a variety of speakers, and on the utterance level to ensure that sufficient speech samples are included for each individual speaker. A significant mismatch in demographic representation between the speaker and utterance level is undesirable. If the representation of a subgroup is higher on the speaker level than the utterance level, this misrepresents the demographics that matter during evaluation and may indicate underrepresentation of individual speakers. Conversely, if the representation of a subgroup is lower on the speaker level, this increases the utterance count per speaker, suggesting overrepresentation of individual speakers. When considering utterances instead of speakers, the representation of females in relation to males decreases from 61\% to 42\% for \emph{VoxCeleb~1 test}, from 82\% to 72\% for \emph{VoxCeleb~1-E} and from 79\% to 70\% for \emph{VoxCeleb~1-H}. The evaluation sets thus not only contain fewer female speakers, they also contain fewer utterances for each female speaker, which reduces the quality of evaluation for female speakers.

\begin{figure}[b]
\vspace{3mm}
    \centering
    \includegraphics[width=\linewidth]{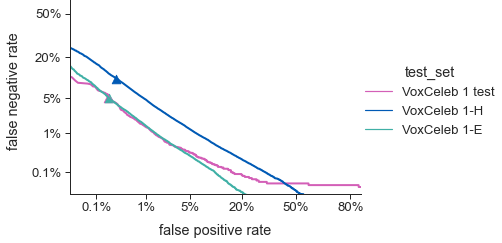}
    \caption{Evaluation bias in the three VoxCeleb~1 evaluation sets with \emph{ResNetSE34V2}}
    \label{fig:det_curves_testsets}
\end{figure}

We evaluate \emph{ResNetSE34V2} with the three evaluation sets and plot the resulting DET curves in Figure~\ref{fig:det_curves_testsets}. The DET curve of \emph{VoxCeleb~1 test} is irregular, confirming that this evaluation set is too small for a valid evaluation. In a FPR range between 0.1\% and 5\%, which is a reasonable operating range for speaker verification, model performance is similar on \emph{VoxCeleb~1 test} and \emph{VoxCeleb~1-E}. The curve of \emph{VoxCeleb~1-H} lies significantly above the other two evaluation sets, indicating that the model performs worse on this evaluation set. Our empirical results illustrate that model performance is highly susceptible to the evaluation set, and show how evaluation bias can affect speaker verification models during evaluation.

\subsubsection{Evaluation Metrics}
The two dominant metrics used in speaker verification benchmarks, including the VoxCeleb SRC, are the equal error rate (EER) and the minimum value of the detection cost function $C_{Det}(\theta_{@\ overall\ min})$ (see Table~\ref{tab:sv_challenges}). Both error metrics give rise to evaluation bias. The EER presents an oversimplified view of model performance, as it cannot weight false positives and false negatives differently. Yet, most speaker verification applications strongly favour either a low FPR or a low FNR~\cite{greenberg2020two}. The NIST SREs do not promote the use of the EER for speaker verification evaluation for this reason \cite{greenberg2020two}, which makes it particularly concerning that new challenges like the SUPERB benchmark evaluate only the EER~\cite{yang2021superb}. $C_{Det}(\theta_{@\ overall\ min})$ can weight FPR and FNR, but has its own shortcomings. Firstly, the detection cost function has been updated over the years, and different versions of the metric are in use. This is impractical for consistent evaluation of applications across time. Secondly, the cost function is only useful if the FPR and FNR weighting reflect the requirements of the application. Determining appropriate weights is a normative design decision, and has received very limited attention in the research community. In benchmarks weights are typically not adjusted, which oversimplifies real-life evaluation scenarios. Finally,  $C_{Det}(\theta_{@\ overall\ min})$ presents a limited view of a model's performance at a single threshold value. While DET curves can provide a holistic view on the performance of speaker verification models across thresholds, many recent research papers do not show them, and those that do only show aggregate curves.

The aggregate form of current evaluation practices based on and optimised for average performance hides the nature of harm that arises from evaluation bias. Ultimately, what matters when a speaker verification system is deployed, are the FPR and FNR. False positives pose a security risk, as they grant unauthorized speakers access to the system. False negatives pose a risk of exclusion, as they deny authorized speakers access to the system. We consider the FPR and FNR for subgroups at $C_{Det}(\theta_{@\ overall\ min})$ in relation to the average FPR and FNR in Table \ref{tab:fpfn_ratios_resnetse34v22} in the Appendix. US male speakers have a FPR and FNR ratio of 1, indicating that this subgroup will experience error rates in line with the average. On the other end of the spectrum Indian female speakers have a FPR and FNR that are 13 and 1.3 times greater than average, indicating that this subgroup is exposed to a significant security risk, and a greater risk of exclusion. 

\subsection{Deployment Bias}
\label{ss:deploymentbias}
\emph{Deployment bias arises when the application context and usage environment do not match the problem space as it was conceptualised during model development.}


\subsubsection{Application Context}
Advancements in speaker verification research have been funded by governments to advance intelligence, defense and justice objectives~\cite{greenberg2020two}. The underlying use cases of speaker verification in these domains have been biometric identification and authentication. From this lens, the speaker verification problem space has been conceptualized to minimize false positives, which result in security breeches. Research on evaluation and consequently also model development has thus focused on attaining low FPRs. This dominant, but limited view promotes deployment bias in new use cases, which require evaluation practices and evaluation datasets tailored to their context.

Today, speaker verification is used in a wide range of audio-based applications, ranging from voice assistants on smart speakers and mobile phones to call centers. A low FPR is necessary to ensure system security. In voice assistants, false positives also affect user privacy, as positive classifications trigger voice data to be sent to service providers for downstream processing~\cite{Schonherr2020Unacceptable}. When used in forensic applications, false positives can amplify existing bias in decision-making systems, for example in the criminal justice system~\cite{machinebias2016}. Even if the FPR is low, the speaker verification system will have a high FNR as trade-off, and the consequences of this must be considered. The FNR affects usability and can lead to a denial of service from voice-based user interfaces. The more critical the service, the higher the risk of harm associated with the FNR. Consider, for example, the previously mentioned speaker verification system used as proof-of-life of pensioners~\cite{veridas2021}. As long as the system is able to identify a pensioner correctly, it relieves the elderly from needing to travel to administrative offices, thus saving them time, money and physical strain. If the system has disparate FNR between demographic subgroups, some populations will be subjected to a greater burden of travel.

Evaluation practices aside, many speaker verification applications will suffer from deployment bias when evaluated on the utterance pairs in the VoxCeleb~1 evaluation datasets. Voice assistants in homes, cars, offices and public spaces are geographically bound, and speakers using them will frequently share a nationality, language and accent. These user and usage contexts should be reflected in the evaluation sets. The VoxCeleb~1 evaluation sets with randomly generated utterance pairs (i.e. \emph{VoxCeleb~1 test} and \emph{-E}) are inadequate to capture speaker verification performance in these application scenarios. Even \emph{VoxCeleb~1-H}, which derives its abbreviation~\emph{-H} from being considered the \emph{hard} evaluation set, is inadequate to evaluate speaker verification performance in very common voice assistant scenarios, such as distinguishing family members. Furthermore, the naming convention of the evaluation sets promotes a limited perspective on speaker verification application contexts: naming \emph{VoxCeleb~1-H} the \emph{hard} evaluation set creates a false impression that the randomly generated utterance pairs of \emph{VoxCeleb~1-E} are the typical evaluation scenario.

\subsubsection{Post-processing}
The operating threshold of a speaker verification system is calibrated after model training (see \S\ref{ss:speaker_verification_evaluation}). This post-processing step amplifies aggregation bias (discussed in \S\ref{ss:aggregationbias}) and deployment bias due to the application context (discussed above). The operating threshold is set in a calibration process that tunes a speaker verification system to a particular evaluation set. If the evaluation set does not take the usage environment and the characteristics of speakers in the environment into consideration, this can give rise to further deployment bias due to post-processing. As discussed above, the VoxCeleb~1 evaluation sets encompass a very limited perspective on application scenarios, and thresholds tuned to these evaluation sets will suffer from deployment bias due to post-processing in many contexts.

The speaker verification system threshold is typically calibrated for the overall evaluation set. This gives rise to a form of aggregation bias that arises during post-processing and deployment. Instead of calibrating the threshold to the overall evaluation set, it could be tuned for each subgroup individually. Using the detection cost function as example, this means setting the threshold for a subgroup to the value $\theta$ where $C_{Det}\left(\theta\right)$ is minimized for the subgroup (i.e. $C_{Det}\left(\theta_{@\ SG\ min}\right)^{SG}$). If the detection cost at the subgroup's minimum is smaller than at the overall minimum, then the subgroup benefits from being tuned to its own threshold. By calculating the ratio of the subgroup's overall detection cost and the subgroup's minimum detection cost, we can get an intuition of the extent of bias. If the ratio is greater than 1, the subgroup will benefit from being tuned to its own threshold. The greater the ratio, the greater the bias and the more the subgroup will benefit from being tuned to its own minimum. Table~\ref{tab:cdet_ratios_resnetse34v22} in the Appendix shows the ratios for all subgroups. It is clear that all subgroups would perform better if tuned to their own threshold. However, female speakers with a mean ratio of 1.37 will experience greater benefit from threshold tuning than male speakers with a mean ratio of 1.09. Visually, the effect of calibrating subgroups to their own threshold can be seen in Figure~\ref{fig:resnetse34v22_det_curves_nationality_} in the Appendix.
\section{Discussion}
\label{s:discussion}

In this paper we have presented an in-depth study of bias in speaker verification, the data processing technique underlying voice biometrics, and a core task in automated speaker recognition. We have provided empirical and analytical evidence of sources of bias at every stage of the speaker verification ML development workflow. Our study highlights that speaker verification performance degradation due to demographic attributes of speakers is significant, and can be attributed to aggregation, learning, evaluation, deployment, historical, representation and measurement bias. Our findings echo concerns similar to those raised in the evaluation for facial recognition technologies~\cite{Raji2021aboutface}. While our findings are specific to speaker verification, they can, for the most part, be extended to automated speaker recognition more broadly. Below we present recommendations for mitigating bias in automated speaker recognition and discuss limitations of our work.




\subsection{Recommendations}

\subsubsection{Inclusive Evaluation Datasets for Real Usage Scenarios.}
We have shown that speaker verification evaluation is extremely sensitive to the evaluation set. The three evaluation sets specified for the VoxCeleb~1 dataset induce evaluation bias, and are insufficient for evaluating many real-world application scenarios. Representative evaluation datasets that are inclusive on a speaker and utterance level are thus needed. On an utterance level, an appropriate evaluation set should contain sufficient utterance pairs for all speakers, and pairs should reflect the reality of the application context. This requires guidelines for constructing application-specific utterance pairs for evaluation. As discussed in~\S\ref{ss:measurement_bias}, our approach for constructing subgroups replicates measurement bias in the labelling choices of VoxCeleb. Future work should consider speaker groups based on vocal characteristics such as pitch, speaking rate, and vocal effort, and consider speaker diversity across languages and accents. Moreover, research on diversity and inclusion in subgroup selection~\cite{Mitchell2020Diversity} presents a starting point that can inform the design of more inclusive speaker verification evaluation datasets.  

\subsubsection{Evaluation Metrics that Consider Consequences of Errors.}
Considering the consequences of errors across application contexts is necessary to reduce deployment bias in speaker verification. Speaker verification evaluation and testing should carefully consider the choice and parameters of error metrics to present robust evaluations and comparison across models for specific application contexts. To this end, guidelines are needed for designing application specific error metrics, and for evaluating bias with these metrics. Such guidelines should determine acceptable FPR and FNR ranges, and guide normative decisions pertaining to the selection of weights of cost functions. Alternative evaluation metrics, such as those used for privacy-preserving speaker verification~\cite{Maouche2020comparative, Nautsch2020privacy}, should also be studied for evaluating bias. To assess aggregation bias in speaker verification, disaggreated evaluation across speaker subgroups is needed. DET curves, which have history in speaker verification evaluation, should be used for visualizing model performance across speaker subgroups. Additionally error metrics should also be computed and compared across subgroups to mitigate evaluation bias.

\subsubsection{Learning and Engineering Approaches for Mitigating Bias.}
Bias in speaker recognition is a new area of study, and interventions are needed to address \emph{learning}, \emph{deployment}, \emph{aggregation} and \emph{measurement bias}. We make some suggestions for interventions that can mitigate these types of bias. Speaker verification will improve for all subgroups if they are tuned to their own threshold rather than the overall threshold. Developing engineering approaches to dynamically select the optimal threshold for subgroups or individual speakers will improve the performance of speaker verification in deployed applications. Subgroup membership is typically not known at run time, making this a challenging task with potential trade-offs against privacy. Further research is also required to study how optimisation for on-device settings, such as model compression, pruning and small-footprint architectures, affect \emph{learning bias}. Previous work in audio keyword spotting has shown that performance disparities across speaker subgroups can be attributed to model input features and the data sample rate at which the voice signal was recorded~\cite{Toussaint2021Characterising}. Studying and mitigating sources of \emph{measurement bias} due to data processing and input features thus remain an important area for future work.


\subsection{Limitations}

Our work presents the first study of bias in speaker verification development and does not study bias in commercial products, which we position as an area for future work. Our aim was to study typical development and evaluation practices in the speaker verification community, not to compare speaker verification algorithms. We thus designed a case study with a confined scope, using publicly available benchmark models as black box predictors. Our findings should be interpreted with this in mind, and not be seen as a generic evaluation for all speaker verification models. We constructed demographic subgroups based on those included in the \emph{VoxCeleb1-H} evaluation set. Some subgroups thus have insufficient sample sizes, which affects the quality of our empirical evaluation for these subgroups. However, as discussed in detail in \S\ref{ss:evaluationbias}, small subgroups are in themselves a source of representation bias that needs to be addressed. We observed that the performance difference that we identified between male and female speakers, and across nationalities, persist across small and large subgroups. 
\section{Conclusion}
\label{s:conclusion}

Automated speaker recognition is deployed on billions of smart devices and in services such as call centres. In this paper we study bias in speaker verification, the biometrics of voice, which is a core task in automated speaker recognition. We present an in-depth empirical and analytical study of bias in a benchmark speaker verification challenge, and show that bias exists at every stage of the machine learning development workflow. Most affected by bias are female speakers and non-US nationalities, who experience significant performance degradation due to aggregation, learning, evaluation, deployment, historic and representation bias. Our findings lay a strong foundation for future work on bias and fairness in automated speaker recognition.

\begin{acks}
This research was partially supported by projects funded by EU Horizon 2020 research and innovation programme under GA No. 101021808 and GA No. 952215.
\end{acks}

\bibliographystyle{ACM-Reference-Format}
\bibliography{references}

\clearpage
\onecolumn
\appendix
\section{Appendix}
\label{appendix}

\subsection{Speaker Verification Evaluation}


\medskip
\textbf{Example Speaker Verification Output Score Distributions}
\begin{figure*}[h!]
    \centering
    \includegraphics[width=0.95\textwidth]{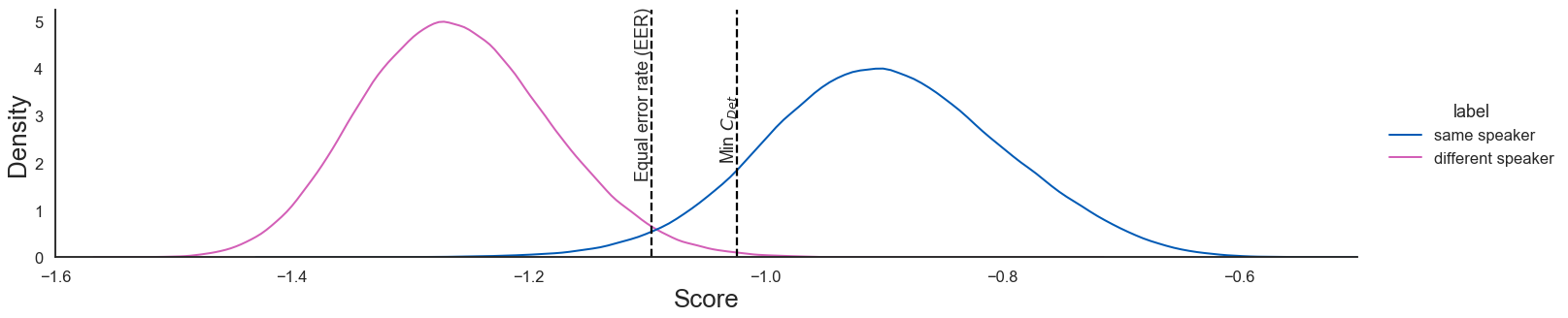}
    \vspace{-2mm}
    \caption{Distribution of speaker verification scores: blue are same speaker trials, pink are different speaker trials. The dotted lines are possible threshold values. Scores to the left of a threshold are rejected, scores to the right are accepted.}
    \label{fig:scores_example}
\end{figure*}

\textbf{Example Detection Error Trade-off Curves}
\begin{figure*}[h!]
    \centering
    \includegraphics[width=0.4\textwidth]{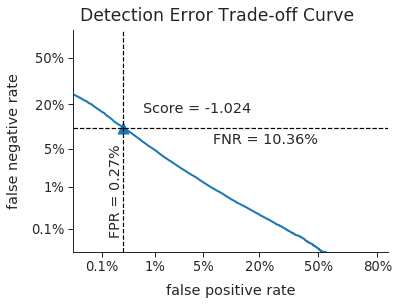}
    \caption{Detection Error Trade-off (DET) curve of a speaker verification system: the blue line shows false positive and false negative error rates at different score values. For example, at the blue triangle the score = -1.024, FPR = 0.27\% and FNR = 10.36\%}
    \label{fig:det_curve_example}
\end{figure*}

\medskip
\textbf{Summary of Technical Details of VoxCeleb SRC Baseline Models}
\begin{table*}[htb]
\small
    \centering
    \begin{tabular}{l|ll}
        Model & \emph{ResNetSE34V2} & \emph{ResNetSE34L} \\ \midrule
        Published in: & \cite{heo2020clova} & \cite{chung2020defence} \\
        Alternative name in publication: & performance optimised model, H/ASP & Fast ResNet-34 \\
        Additional training procedures: & data augmentation (noise \& room impulse response) & - \\       
        Parameters: & 8 million & 1.4 million\\       
        Frame-level aggregation: & attentive statistical pooling & self-attentive pooling\\       
        Loss function: & angular portotypical softmax loss & angular portotypical loss\\       
        Input features: & 64 dim log Mel filterbanks & 40 dim Mel filterbanks\\
        Window (width x step): & 25ms x 10ms & 25ms x 10ms\\
        Optimized for: & predictive performance & fast execution \\
    \end{tabular}
    \medskip
    \caption{Attributes of two VoxCeleb SRC baseline models} 
    \label{tab:baseline_model_attributes}
\end{table*}

\clearpage

\subsection{Representation Bias}
\begin{figure*}[hbt]
    \centering
    \includegraphics[width=0.9\textwidth]{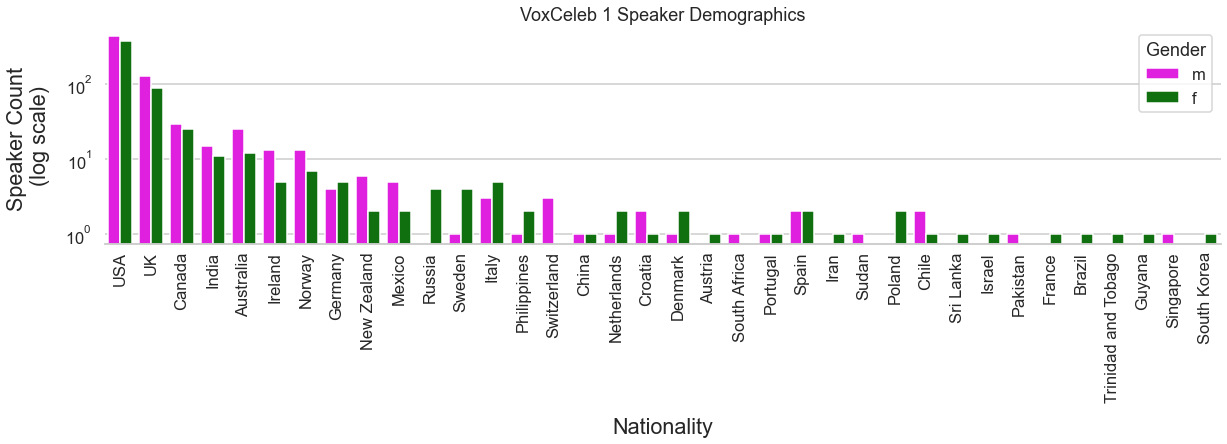}
    \vspace{-5mm}
    \caption{VoxCeleb 1 Speaker Demographics}
    \label{fig:speaker_demographics}
    \vspace{-5mm}
\end{figure*}

\clearpage
\subsection{Evaluation Bias}

\begin{figure*}[hbt]
    \centering
    \includegraphics[width=\textwidth]{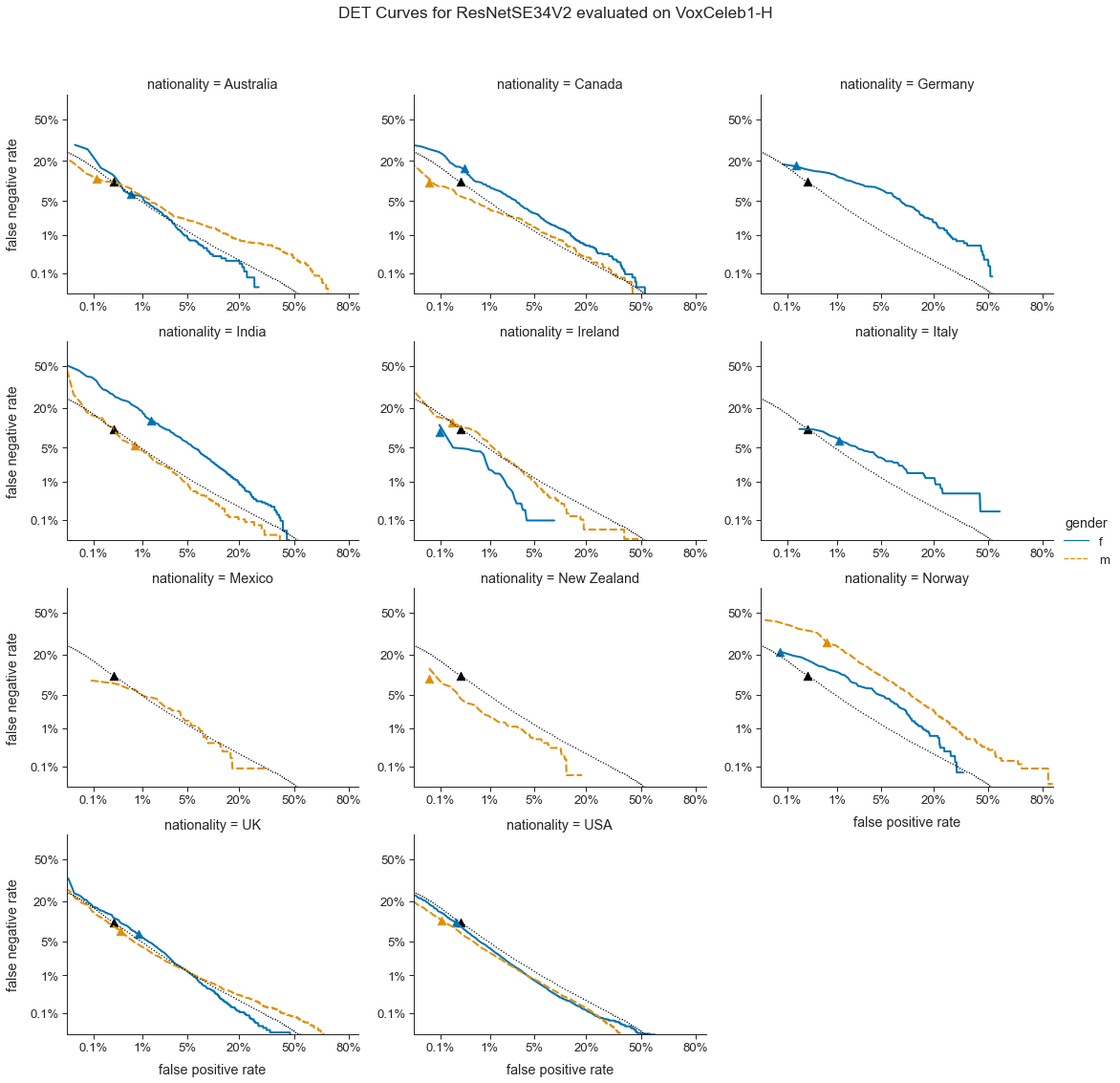}
    \caption{ \emph{ResNetSE34V2} DET curves for speaker subgroups evaluated on the \emph{VoxCeleb~1-H} evaluation set. The dotted black lines indicate the aggregate overall DET curve across all subgroups.}
    \label{fig:resnetse34v22_det_curves_nationality_all}
\end{figure*}
\clearpage

\subsection{Learning Bias}

\begin{figure*}[hb]
    \centering
    \includegraphics[width=\textwidth]{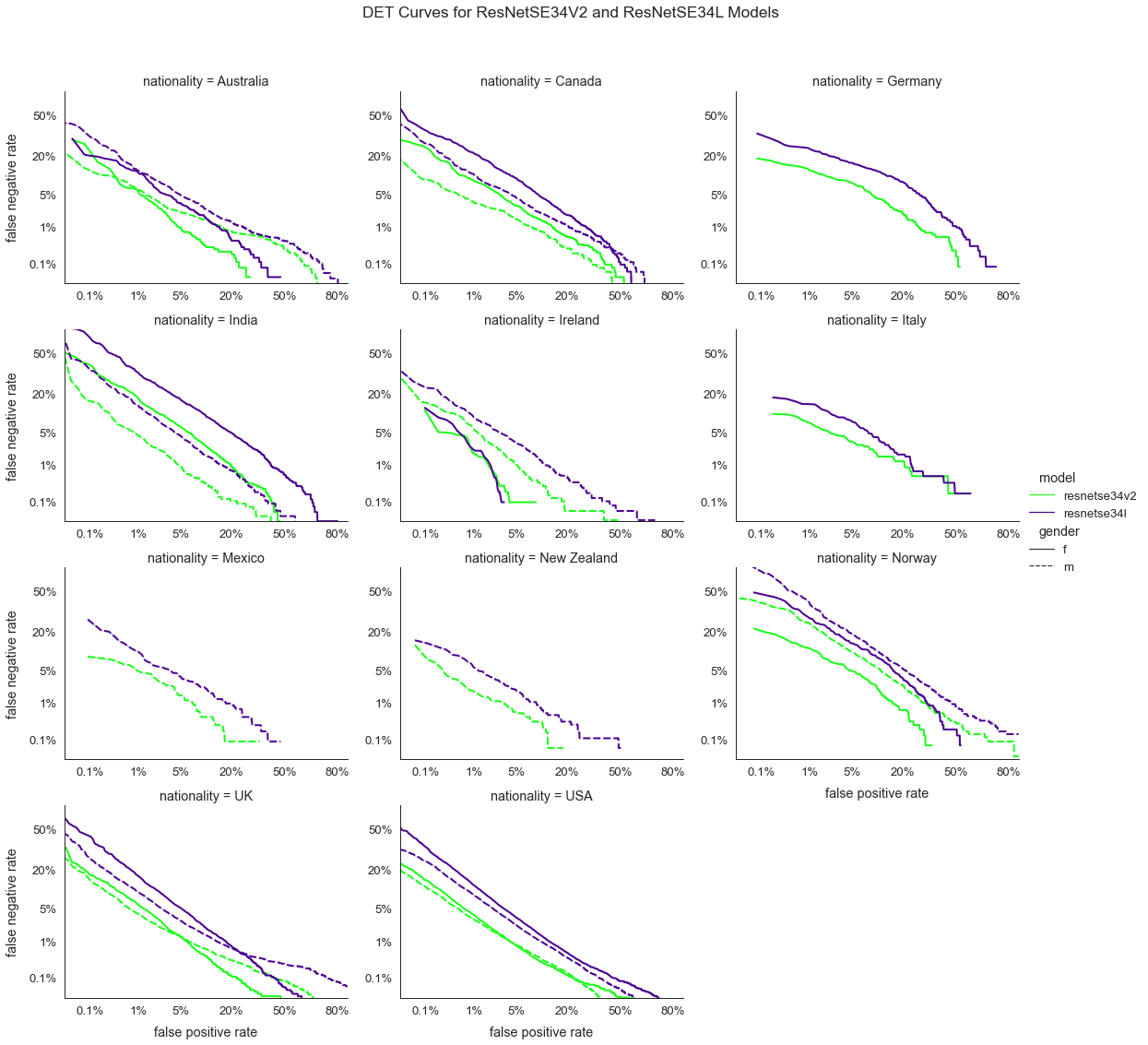}
    \caption{Learning bias based on model architecture}
    \label{fig:det_curves_models_nationality}
\end{figure*}
\clearpage


\subsection{Aggregation and Post-processing Deployment Bias}
\begin{table}[hbt]
\small
\centering
\begin{tabular}{lccccc}
\multicolumn{1}{l}{\textbf{Subgroup (SG)}} &
  \textbf{\begin{tabular}[c]{@{}c@{}}Unique \\ speakers\end{tabular}} &
  \textbf{\begin{tabular}[c]{@{}c@{}}$C_{Det}(\theta_{@\ overall\ min})^{SG}$\end{tabular}} &
  \textbf{$subgroup\ bias$} &
  \textbf{\begin{tabular}[c]{@{}c@{}}$C_{Det}(\theta_{@\ SG\ min})^{SG}$\end{tabular}} &
  \textbf{\begin{tabular}[c]{@{}c@{}}$threshold\ bias$\end{tabular}} \\ \noalign{\smallskip}
\textbf{mexico\_m} & 5 & 0.090 & 0.5768 & 0.090 & 1.0000 \\
\textbf{newzealand\_m} & 6 & 0.104 & 0.6668 & 0.086 & 1.2093 \\
\textbf{ireland\_f} & 5 & 0.110 & 0.7109 & 0.070 & 1.5714 \\
\textbf{canada\_m} & 29 & 0.114 & 0.7304 & 0.104 & 1.0962 \\
\textbf{usa\_m} & 431 & 0.130 & 0.8357 & 0.122 & 1.0656 \\
\textbf{australia\_m} & 25 & 0.140 & 0.9020 & 0.136 & 1.0294 \\
\textbf{usa\_f} & 368 & 0.142 & 0.9224 & 0.140 & 1.0143 \\
\textbf{uk\_m} & 127 & 0.148 & 0.9523 & 0.140 & 1.0571 \\ \cline{1-1} \cline{4-4}
\textbf{ireland\_m} & 13 & 0.162 & 1.0432 & 0.160 & 1.0125 \\
\textbf{australia\_f} & 12 & 0.178 & 1.1523 & 0.154 & 1.1558 \\
\textbf{india\_m} & 15 & 0.190 & 1.2200 & 0.144 & 1.3194 \\
\textbf{germany\_f} & 5 & 0.208 & 1.3359 & 0.184 & 1.1304 \\
\textbf{canada\_f} & 25 & 0.224 & 1.4501 & 0.202 & 1.1089 \\
\textbf{uk\_f} & 88 & 0.226 & 1.4558 & 0.172 & 1.3140 \\
\textbf{norway\_f} & 7 & 0.228 & 1.4711 & 0.210 & 1.0857 \\
\textbf{italy\_f} & 5 & 0.276 & 1.7827 & 0.104 & 2.6538 \\
\textbf{norway\_m} & 13 & 0.398 & 2.5720 & 0.396 & 1.0051 \\
\textbf{india\_f} & 11 & 0.400 & 2.5766 & 0.318 & 1.2579
\end{tabular} {\medskip}
\caption{Detection costs, \emph{subgroup bias} and post-processing aggregation bias (see Equation~\ref{eq:thresholdbias}) for subgroups at overall and subgroup minimum thresholds with $C_{Det}(\theta_{@\ overall\ min})^{overall} = 0.154$. Subgroups above the horizontal black line have a \emph{subgroup bias} less than 1 and perform better than average when tuned to $C_{Det}(\theta_{@\ overall\ min})$. Female subgroups are on average subjected to more bias than male subgroups.}
\label{tab:cdet_ratios_resnetse34v22}
\vspace{-10mm}
\end{table}

\begin{equation}
\label{eq:thresholdbias}
    threshold\ bias = \frac{C_{Det}\left(\theta_{@\ overall\ min}\right)^{SG}}{C_{Det}\left(\theta_{@\ SG\ min}\right)^{SG}}
    \vspace{3mm}
\end{equation}

\begin{figure*}[htb]
    \centering
    \vspace{5mm}
    \includegraphics[width=\textwidth]{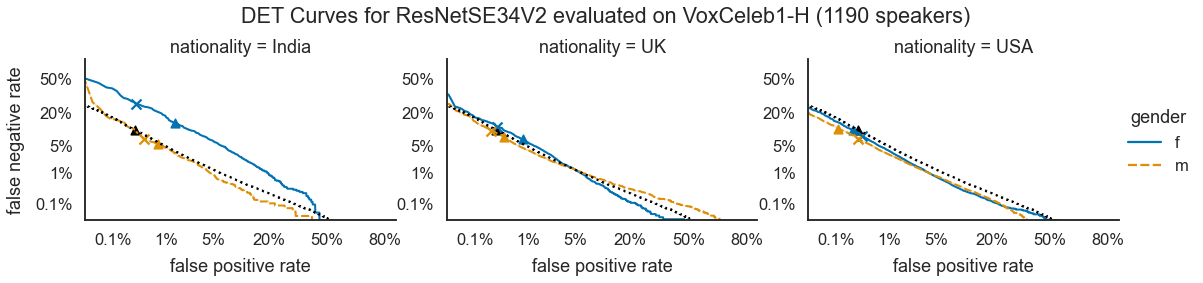}
    \caption{DET curves and thresholds for male and female speakers of Indian, UK and USA nationalities for ResNetSE34V2 evaluated on the VoxCeleb1-H test set.}
    \label{fig:resnetse34v22_det_curves_nationality_}
\end{figure*}

\noindent
Figure~\ref{fig:resnetse34v22_det_curves_nationality_} shows DET curves and thresholds for ResNetSE34V2 for male and female speakers of Indian, UK and USA nationalities evaluated on \emph{VoxCeleb~1-H}. We use the following conventions: triangle markers show the FPR and FNR at the overall minimum threshold $C_{Det}\left(\theta_{@\ overall\ min}\right)^{SG}$, cross markers show the FPR and FNR at the subgroup minimum threshold $C_{Det}\left(\theta_{@\ SG\ min}\right)$, and dotted black lines and markers are used for the overall DET curve and threshold. The DET curve of female Indian speakers lies far above the overall aggregate, indicating that irrespective of the threshold, the model will always perform worse than aggregate for this subgroup. In the operating region around the tuned thresholds, the model also performs worse for female speakers from both the UK and the USA. Being tuned to $C_{Det}\left(\theta_{@\ overall\ min}\right)$ does not affect the FNR and improves the FPR of USA female and male speakers. For other speaker subgroups, especially UK females and Indian females and males, either the FPR or the FNR deteriorates significantly when tuned to the overall minimum. For all subgroups the threshold at the subgroup minimum, $C_{Det}\left(\theta_{@\ SG\ min}\right)$, shifts the FPR and FNR closer to those of the minimum overall threshold, suggesting that performance will improve when optimising thresholds for subgroups individually.

\subsection{Application Context Deployment Bias}

\begin{SCtable}[30][htb]
\small
\centering
\begin{tabular}{lccc}
\textbf{Subgroup} &
  \textbf{\begin{tabular}[c]{@{}c@{}}Unique \\ speakers\end{tabular}} &
  \textbf{\begin{tabular}[c]{@{}c@{}}FPR ratio \\ overall\end{tabular}} &
  \textbf{\begin{tabular}[c]{@{}c@{}}FNR ratio \\ overall\end{tabular}} \\ \noalign{\smallskip}
\textbf{mexico\_m}     & 5   & 0.0000  & 0.8173 \\
\textbf{canada\_m}     & 29  & 0.5171  & 0.9396 \\
\textbf{newzealand\_m} & 6   & 0.5218  & 0.8487 \\
\textbf{norway\_f}     & 7   & 0.6306  & 1.9682 \\
\textbf{ireland\_f}    & 5   & 0.9037  & 0.8408 \\
\textbf{usa\_m}        & 431 & 1.0000  & 1.0000 \\
\textbf{australia\_m}  & 25  & 1.1055  & 1.0745 \\
\textbf{germany\_f}    & 5   & 1.5023  & 1.6162 \\
\textbf{ireland\_m}    & 13  & 1.6864  & 1.1675 \\
\textbf{usa\_f}        & 368 & 2.0542  & 0.9287 \\
\textbf{canada\_f}     & 25  & 3.1483  & 1.4749 \\
\textbf{uk\_m}         & 127 & 3.5339  & 0.6986 \\
\textbf{australia\_f}  & 12  & 5.6031  & 0.6008 \\
\textbf{norway\_m}     & 13  & 6.0866  & 2.5233 \\
\textbf{india\_m}      & 15  & 6.6852  & 0.4975 \\
\textbf{uk\_f}         & 88  & 7.8514  & 0.6168 \\
\textbf{italy\_f}      & 5   & 10.3484 & 0.6202 \\
\textbf{india\_f}      & 11  & 13.0387 & 1.2497
\end{tabular}
\caption{FPR and FNR ratios for subgroups at $C_{Det}(\theta_{@\ overall\ min})$. The ratio is calculated by dividing the subgroup FPR and FNR by the overall FPR and FNR respectively. It thus presents a relative view on how much better or worse the subgroup error rates are in relation to the overall error rates.}
\label{tab:fpfn_ratios_resnetse34v22}
\end{SCtable}


\end{document}